\documentclass[usenatbib,twocolumn]{article}
\usepackage[a4paper, margin=0.5in]{geometry}

\usepackage{graphics}
\usepackage{graphicx}
\usepackage{color}
\usepackage[english]{babel}
\usepackage{aas_macros}
\usepackage{natbib}
\usepackage{amsmath}
\usepackage{amssymb}
\usepackage{comment}
\usepackage{soul} % strike through
\usepackage{gensymb}
\usepackage{subcaption}
\usepackage{flushend}
\usepackage{url}
\title{Compactified Cosmological Simulations of the Infinite Universe}

\begin{document}

\author{G\'abor R\'acz$^1$\thanks{E-mail: ragraat@caesar.elte.hu},  Istv\'an Szapudi$^2$, Istv\'an Csabai$^1$, L\'aszl\'o Dobos$^1$
\\
$^1$Department of Physics of Complex Systems, E\"{o}tv\"{o}s Lor\'and University, Pf. 32, H-1518 Budapest, Hungary\\
$^2$Institute for Astronomy, University of Hawaii, 2680 Woodlawn Drive, Honolulu, HI, 96822
}

\maketitle

\begin{abstract}

We present a novel $N$-body simulation method that compactifies the infinite spatial extent of the Universe into a finite sphere with isotropic boundary conditions to follow the evolution of the large-scale structure. Our approach eliminates the need for periodic boundary conditions, a mere numerical convenience which is not supported by observation and which modifies the law of force on large scales in an unrealistic fashion. We demonstrate that our method outclasses standard simulations executed on workstation-scale hardware in dynamic range, it is balanced in following a comparable number of high and low $k$ modes and, its fundamental geometry and topology match observations. Our approach is also capable of simulating an expanding, infinite universe in static coordinates with Newtonian dynamics. The price of these achievements is that most of the simulated volume has smoothly varying mass and spatial resolution, an approximation that carries different systematics than periodic simulations.

Our initial implementation of the method is called StePS which stands for Stereographically Projected Cosmological Simulations. It uses stereographic projection for space compactification and naive $\mathcal{O}(N^2)$ force calculation which is nevertheless faster to arrive at a correlation function of the same quality than any standard (tree or P$^3$M) algorithm with similar spatial and mass resolution. The $N^2$ force calculation is easy to adapt to modern graphics cards, hence our code can function as a high-speed prediction tool for modern large-scale surveys. To learn about the limits of the respective methods, we compare StePS with GADGET-2 
%\citep{Gadget2_2005MNRAS.364.1105S} 
running matching initial conditions.

%Periodic boundary conditions are equivalent to the assumption of a toroidal topology for the Universe that is not supported by observations. 

\end{abstract}

\section{Introduction}

Cosmological $N$-body simulations are commonly used to calculate the non-linear evolution of dark matter at late times. Although multi-resolution algorithms, e.g. adaptive refinement \citep{1999ASSL..240...19N}, exist, standard simulations usually have constant mass and spatial resolution throughout the simulation volume and assume periodic or quasi-periodic boundary conditions to account for the effect of the infinite volume outside the simulation box. This assumption is only a matter of convenience as it contradicts fundamental facts. Observations do not support the compact toroidal topology of space, nor the large-scale modifications to the law of force these simulations carry \citep{2016Univ....2....1L, 1995PhR...254..135L}. Although this latter effect can be accounted for and mitigated by executing large-volume simulations, it is only possible at the cost of high-performance computing. Moreover, constant mass and spatial resolution simulations are inefficient in the sense that they follow an exorbitant number of high $k$ modes and at the same time, since the number of modes grows as $k^3$, low $k$ modes are miserably undersampled, even in the largest simulation boxes.

The standard solutions to the problem of low $k$ modes are the somewhat arbitrary and bespoke zoom-in sequence simulations and adaptive refinement schemes, in which the algorithm decides where the higher resolution is needed. Zoom-in simulations are usually executed in two steps \citep{1994MNRAS.267..401N}. First, to minimize the effect of periodic boundary conditions and to follow a large enough number of low $k$ modes, a parent simulation is run at low mass resolution but in a large simulation volume. Then a small ``volume of interest'' is chosen and re-simulated at high resolution, while the low resolution simulation is replayed in the background to provide the boundary conditions. In some schemes, there are more than two levels of simulations nested recursively in order to achieve even higher resolutions \citep{2014MNRAS.437.1894O}.

Finite volume simulations with toroidal topology cannot account for the expansion of the Universe unless coordinates and velocities are rescaled into comoving coordinates according to Friedmann equations. This prevents investigating interesting open questions such as the existence of Newtonian backreaction \citep{2017arXiv170308809K}.

The principal goal of this paper is to develop an algorithm that runs simulations with realistic geometry and topology while also yields a better balance between high and low $k$ modes of the power spectrum. These objectives are achieved by applying a compactification map to the infinite universe to render it into a finite volume. In particular, we present an implementation that uses stereographic projection to map the infinite three-dimensional hyperplane of space onto the surface of a compact four-dimensional sphere. Initial conditions are defined with the help of an equal volume grid on this compact surface. Note, that with any similar compactification map there will be a grid element which corresponds to infinite volume in the original Euclidean space with $\delta = 0$ density fluctuations. When using comoving coordinates, the force produced by this grid element is non-trivial and will have to be substituted with boundary conditions. In the non-comoving Newtonian case, although conceptually we simulate an infinite universe, when $\Lambda = 0$ this grid element can be safely omitted while the remaining, finite volume of the simulation will develop forward in time.

The paper is organized as follows. In Sec.~\ref{sec:eqmotion}, we recap on the equations of motion in comoving coordinates then, in Sec.~\ref{sec:initial}, we discuss the stereographic projection as a possible way to compactify space, calculate the boundary conditions and describe our solution to generating the initial conditions. for compactified simulations. Furthermore, in Sec.~\ref{sec:static}, we outline how compactified simulations could be executed in static coordinates. In Sec.~\ref{sec:angpowspec}, we discuss the angular power spectrum as the natural tool for comparing results from compactified simulations to observations. The results from a StePS simulation are compared to a standard periodic simulation in Sec.~\ref{sec:results} and a summary of our findings and future work directions is given in Sec.~\ref{sec:summary}. \\

The simulation and initial condition generation code is available for download from the paper's web site\footnote{\url{http://www.vo.elte.hu/papers/2017/steps/}}.

\section{Equations of motion with periodic boundary conditions}
\label{sec:eqmotion}

In the low velocity limit, dynamics of dark matter is governed by Newtonian gravity \citep{1980lssu.book.....P, 2017arXiv170308809K}. To simulate an \textit{infinite} universe in a computer with \textit{finite} memory, one can either introduce periodic boundary conditions, which result in a non-physical force law at large scales \citep{1988csup.book.....H, Gadget2_2005MNRAS.364.1105S} or, attempt to increase the simulated volume as much as possible and use isotropic boundary conditions as far from the interesting part of the simulation as possible. In the following, we develop the concept of the latter and determine the isotropic boundary conditions for simulations in static and comoving coordinates.

%In the most common Dark Matter only implemention, the Friedmann equations with the cosmological parameters $\Omega$ and $H_0$  determine the overall expansion rate at each time step, while particles interact via Newtonian dynamics within this expanding metric \citep[e.g.,][]{1988csup.book.....H, Gadget2_2005MNRAS.364.1105S}. Such algorithms provide an adequate approximation, since the dark matter only interacts gravitationally in most models, and the velocities are small compared to the speed of light \citep[e.g.,][]{1980lssu.book.....P, 2017arXiv170308809K}.

Cosmological simulations with periodic boundary conditions account for the expansion of the Universe by rescaling the metric inside the simulation box with a scale factor which can be determined from Friedmann equations. The conventional equations of motion of dark matter particles in comoving coordinates, as used by most standard cosmological $N$-body simulations, are
\begin{equation}
	m_i\cdot\ddot{\mathbf{x}}_i = \sum\limits_{j=1; j \neq i}^{N} \frac{m_im_j\mathbf{F}(\mathbf{x}_i-\mathbf{x}_j, h_i+h_j)}{a(t)^{3}} - 2 \cdot m_i \frac{\dot{a}(t)}{a(t)} \cdot \dot{\mathbf{x}}_i,
\label{eq:Comoving_newtonian}
\end{equation}
where $\mathbf{x}_i$ and $m_i$ are the comoving coordinates and the masses of the particles, respectively, while $a(t)$ is the time-dependent scale factor of the expanding background universe.

\begin{figure}
\centering
\includegraphics[width=\columnwidth]{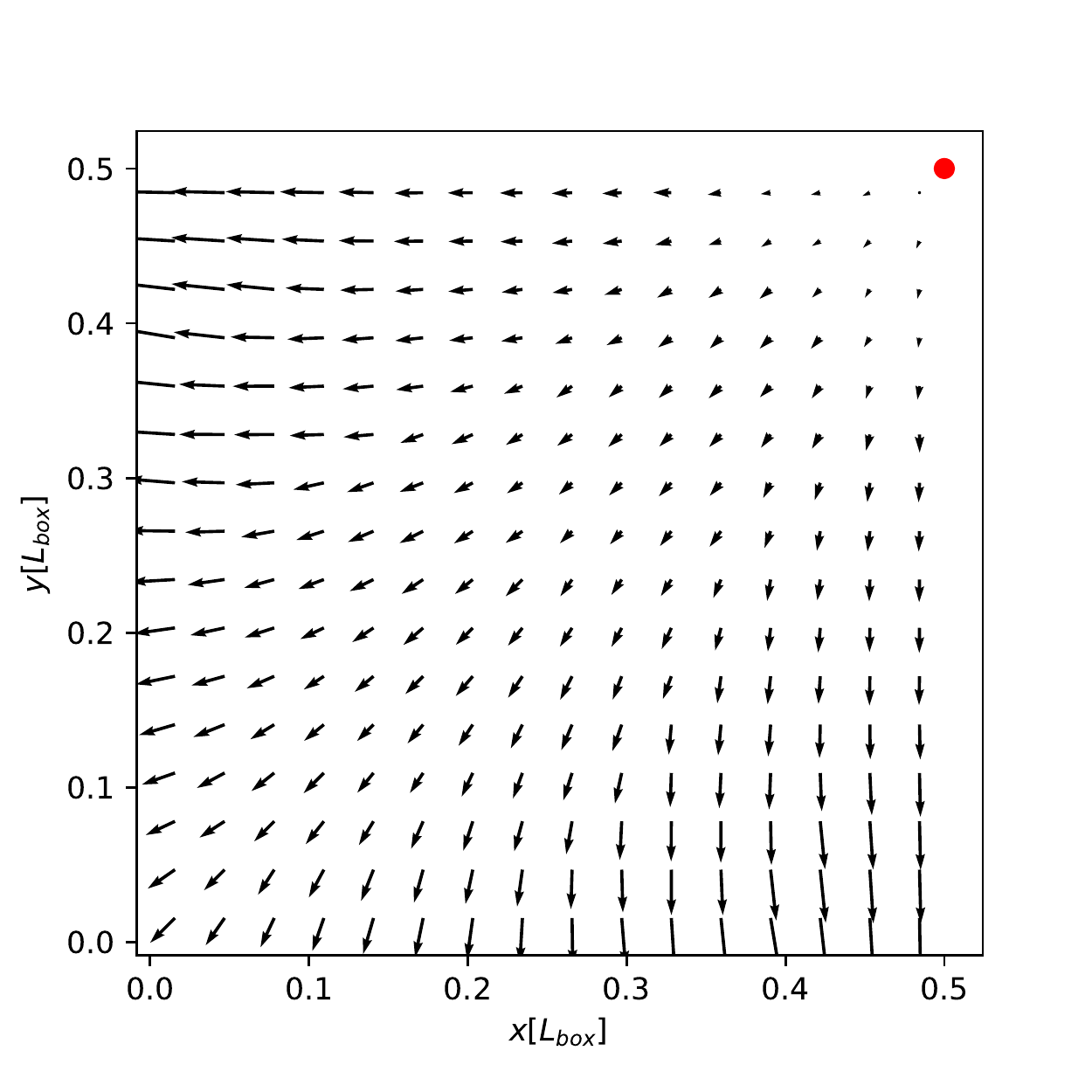}
\caption{The difference $\mathbf{F}_\textnormal{periodic}-\mathbf{F}_\textnormal{infinite}$ of the force fields generated by an isolated mass particle as calculated with and without periodic boundary conditions. The mass particle (red dot) is located at the centre of the simulation box with a linear size of $L_\textnormal{box}$, but we only plotted the bottom left quarter of the box. Ewald summation was used to calculate the force field with periodic boundary conditions. The directions of the arrows are parallel to the direction, the length of the arrows are proportional to the magnitude of the difference between the force fields.}
\label{fig:ForceFields}
\end{figure}

The function $\mathbf{F}(\mathbf{x}_i-\mathbf{x}_j, h_i+h_j)$ is proportional to the gravitational force between particles $i$ and $j$ and depends on the boundary conditions, as well as on $h_i$ and $h_j$, the softening lengths associated with the particles. The mass-independent spline kernel \citep{1985A&A...149..135M, Gadget2_2005MNRAS.364.1105S} is widely used to soften the gravitational force. 
%\begin{equation}
%        \mathcal{F}(r, h) = \left\{
%                \begin{array}{l l}
%                \frac{32{r}^{4}}{{(2h)}^{6}}-\frac{38.4{r}^{3}}{{(2h)}^{5}}+\frac{32r}{3{(2h)}^{3}} & \text{\small{if $r < h$}}\\
%                 \ &\ \\
%		 -\frac{32{r}^{4}}{3{(2h)}^{6}}+\frac{38.4{r}^{3}}{{(2h)}^{5}}-\frac{48{r}^{2}}{{(2h)}^{4}}+\frac{64r}{3\,{(2h)}^{3}}-\frac{1}{15{r}^{2}} & \text{\small{if  $h<r<2h$}}\\
%               \ &\ \\
%                \frac{1}{{r}^{2}} & \text{\small{if $2h<r$}}
%                \end{array} \right. .
%\label{eq:Force_spline_kernel}
%\end{equation}
In case of zero boundary conditions the force takes the form of
\begin{equation}
	\mathbf{F}(\mathbf{x}, h) = -G \cdot \mathcal{F}(\left| \mathbf{x} \right|, h) \cdot \frac{\mathbf{x}}{ \left|\mathbf{x} \right| },
\label{eq:Force}
\end{equation}
where $G$ is the gravitational constant, $h$ is the softening length and $\mathcal{F}(r, h)$ is the softened scalar-valued force function. In case of periodic boundary conditions multiple images of the particles have to be taken into account, hence forces are calculated using Ewald summation as
\begin{equation}
	\mathbf{F}(\mathbf{x}, h) = \sum\limits_{\mathbf{n}}-G \cdot \mathcal{F}( \left| \mathbf{x} - \mathbf{n}L \right|, h) \cdot \frac{\mathbf{x}-\mathbf{n}L}{ \left| \mathbf{x}-\mathbf{n}L \right|},
\label{eq:ForcePeriodic}
\end{equation}
where $L$ is the linear size of the simulation box and $\mathbf{n}=(n_1,n_2,n_3)$ extends over all integer triplets. Obviously, a numerical code cannot sum for all integer triplets, so a cut in $\mathbf{n}$ is necessary. The common choice is where the only valid triplets are $\left| \mathbf{x}-\mathbf{n}L \right| < 2.6L$ \citep{1991ApJS...75..231H}. At large scales, this force law clearly differs from the Newtonian inverse-square law.  To emphasize the effect of periodic boundary conditions on the force law, in Fig.~\ref{fig:ForceFields}, we plot the difference between Eqs.~\ref{eq:Force}~and~\ref{eq:ForcePeriodic} for the case of a single isolated point mass.

\section{Initial and boundary conditions for compactified simulations}
\label{sec:initial}

By applying a spherically symmetric transformation, one can compactify the infinite three-dimensional space into a finite volume. A large class of transformations exists which, at least from an arbitrarily chosen origin, preserve the isotropy of the original space. From now on, we will refer to the original, infinite, three-dimensional space with the Euclidean metric as $\mathcal{P}$. The compactification methods will transform $\mathcal{P}$ onto the surface of a four-dimensional hypersphere $\mathcal{S}$ of radius $R_S$. $R_S$ is not to be confused with $R_\textnormal{sim}$, the radius of the compactified simulation in $\mathcal{P}$ that we will discuss later.

It is easy to construct an equal volume regular grid in the compactified space $\mathcal{S}$ which, when spherical coordinates are used, corresponds to a grid in the original space that is regular in the angular coordinates but grows smoothly in size radially. With the help of such a grid, by averaging the positions and summing up the inertia of the particles in each grid element of the non-uniform grid of the infinite space $\mathcal{P}$, one can easily compactify a particle distribution. %and transform the velocities and masses of the particles into the compactified space $\mathcal{S}$.

\subsection{Stereographic projection in three dimensions}
\label{sec:stereo}

\begin{figure}
\begin{center}
\includegraphics{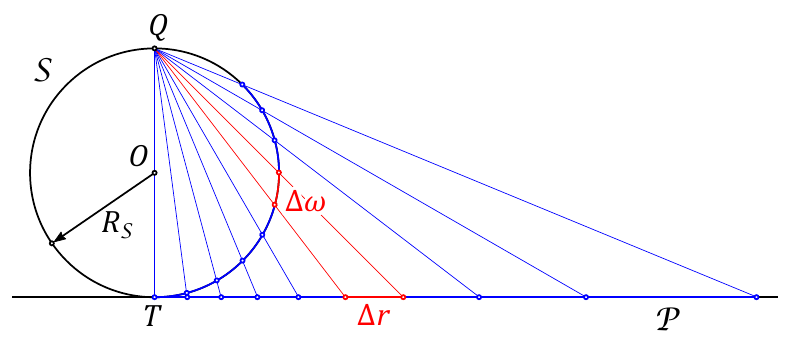}
\end{center}
\vspace*{-0.2cm}
\caption{One-dimensional stereographic projection. To compactify the straight line $\mathcal{P}$ onto the sphere $\mathcal{S}$ touching it at the tangent point $T$, rays from the projection point $Q$ are drawn to the points of $\mathcal{P}$. Each pair of adjacent rays cut out the same arc of $\Delta\omega$ from $\mathcal{S}$ whereas the corresponding length of $\Delta r$ increases drastically with the distance from $T$ on $\mathcal{P}$. The transformation is not defined in $Q$. Any infinitesimally small arc that contains $Q$ is projected into two infinitely long sections of $\mathcal{P}$.}
\label{fig:1DsterProj}
\end{figure}

For our particular implementation of a compactified cosmological simulation, we start from the stereographic projection which is a well-known bijective geometrical transformation that projects the sphere $\mathcal{S}$ onto a plane $\mathcal{P}$. The stereographic projection is straightforward to invert and generalize to three dimensions. In the one-dimensional case, see Fig.~\ref{fig:1DsterProj}, it is a transformation between a circle and one of its tangents with the tangent point $T$. If a point on the circle is parametrised by the length of the arc $\omega$, which is measured toward the given point from $T$, then one finds its stereographic projection as
\begin{equation}
	r = 2 R_S \cdot \tan\left( \frac{\omega}{2} \right),
\label{eq:1Dster_proj}
\end{equation}
where $R_S$ is the radius of the circle and $r$ is the distance of the projection from $T$. The inverse transformation is simply 
\begin{equation}
        \omega = 2 \cdot \arctan\left( \frac{r}{2 R_S} \right).
\label{eq:1Dster_proj_inv}
\end{equation}

\begin{figure}
\begin{center}
\vspace*{-0.5cm}
\includegraphics{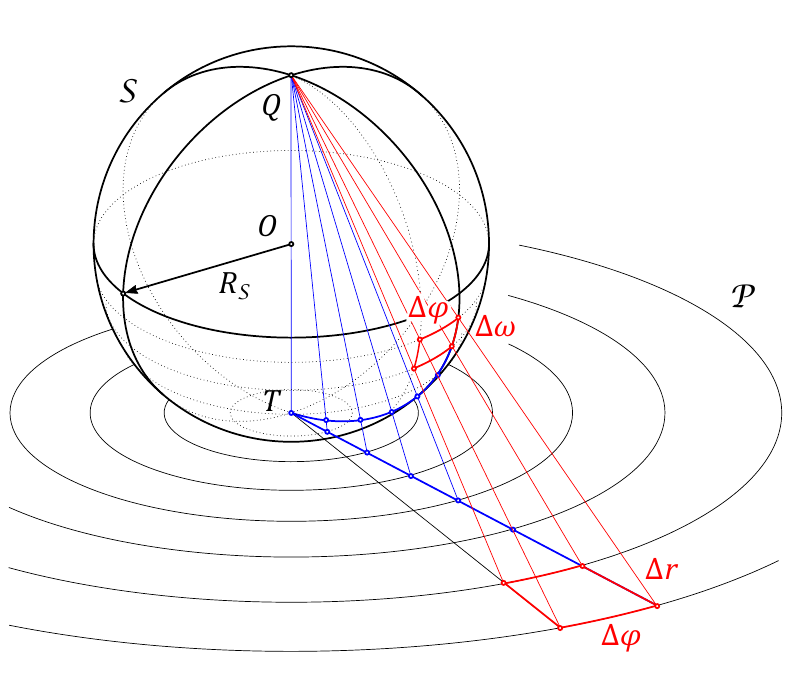}
\end{center}
\vspace*{-0.2cm}
\caption{Two-dimensional stereographic projection that maps the sphere $\mathcal{S}$ to a tangent plane $\mathcal{P}$ and vice versa. Similarly to the one-dimensional case, the equal-length $\Delta\omega$ arcs are projected into sections with increasing length as a function of distance from $T$. At the time, the $\Delta\varphi$ angle is not transformed. Any infinitesimally small circle around $Q$ is projected into the entire plane all the way to infinity minus a circle if finite radius centred on $T$. The stereographic transformation can be trivially generalized into higher dimensions by transforming the first polar angle only (in our case, $\omega$), and keeping all other polar angles and the azimuth angle $\varphi$ identical.}
\label{fig:2DsterProj}
\end{figure}

In two dimensions, the transformation becomes very simple if the tangent plane is parametrised by polar coordinates centred on the tangent point. In this case only the radial coordinate is transformed, the polar angle in the plane remains equal to the azimuthal angle on the sphere, see Fig.~\ref{fig:2DsterProj}.

The three-dimensional stereographic transformation maps between the surface of a four-dimensional sphere and a tangent three-dimensional hyperplane. Let $\omega$, $\vartheta$, and $\varphi$ be the hyperspherical coordinates of the four-dimensional sphere where $\omega, \vartheta \in \left[ 0, \pi \right]$, $\varphi \in \left[0, 2\pi \right]$ and $\omega$ is measured from the tangent point toward the opposite pole $Q$ of the hypersphere. If the three-dimensional hyperplane is parametrised by spherical coordinates centred on the tangent point, only the angle $\omega$ is transformed, and the transformation rules are the same as in Eqs.~\ref{eq:1Dster_proj}~and~\ref{eq:1Dster_proj_inv}, except that $r$ is now the three-dimensional Euclidean distance from the tangent point. The angles $\vartheta$ and $\varphi$ will be identical to the polar and azimuthal angles of the three-dimensional spherical coordinates. The stereographic projection is an obvious compactification of the infinite space in the radial direction: the coordinate $r \in \left[ 0, \infty \right)$ is mapped to the $\omega \in \left[ 0, \pi \right)$ finite interval. When physical units are used, the radius $R_S$ of the four-dimensional sphere that defines the projection will determine the length scale on which the transformation is close to linear.

When Cartesian coordinates are used in the three-dimensional space, transformation rules become 
\begin{equation}
	\begin{aligned}
        x &= 2R_s \cdot \tan\left( \frac{\omega}{2} \right)\sin(\vartheta)\cos(\varphi)\\
	y &= 2R_s \cdot \tan\left( \frac{\omega}{2} \right)\sin(\vartheta)\sin(\varphi)\\
	z &= 2R_s \cdot \tan\left( \frac{\omega}{2} \right)\cos(\vartheta)
	\end{aligned},
\label{eq:3Dster_proj}
\end{equation}
whereas the inverse rules are
\begin{equation}
	\begin{aligned}
        \omega &= 2 \cdot \arctan\left( \frac{\sqrt{x^2+y^2+z^2}}{2R_s} \right)\\
        \vartheta &= \cos^{-1}\left( \frac{z}{\sqrt{x^2+y^2+z^2}} \right)\\
        \varphi &= \arctan\left( \frac{y}{x} \right).
	\end{aligned}
\label{eq:3Dster_proj_inv}
\end{equation}

By tenacious calculations, one can also derive the expressions for the gravitational force in compactified coordinates but the complexity of the formulae would make it inefficient to use them on the computer. Instead, we will determine the initial conditions in the compactified space and project them back to the three-dimensional space to follow the evolution of structure in Cartesian coordinates.

We illustrate the effect of the stereographic projection on equal height rings around the sphere $S$ in Fig.~\ref{fig:1Dmap}. The projection maps the pole $Q$ opposite to the tangent point $T$ to infinity and the small red circle encompassing $Q$ is mapped into an infinite volume outside a finite sphere in $\mathcal{P}$. When executing compactified simulations, we will exclude the small circle from $\mathcal{S}$ which is equivalent to restricting the simulation volume to the red sphere in $\mathcal{P}$ and, as we will show in Sec.~\ref{sec:boundary}, substituting the force originating from the particles in the red circle with isotropic boundary conditions.

\subsection{Compactified initial conditions}
\label{sec:initcond}

Several methods exist to generate periodic initial condition for cosmological N-body simulations. Most of these methods use post-linear theory and apply ``tricks'' to produce a relaxed distribution of point masses with a given correlation function. In case of compactified simulations, it would be favourable to generate initial conditions directly in compactified coordinates but since no method is currently available to do so, we decided to start from high resolution, large volume periodic initial conditions and apply the aforementioned binning technique to transform the particles into the compactified space.

\begin{figure}
\begin{center}
\vspace*{-0.5cm}
\includegraphics{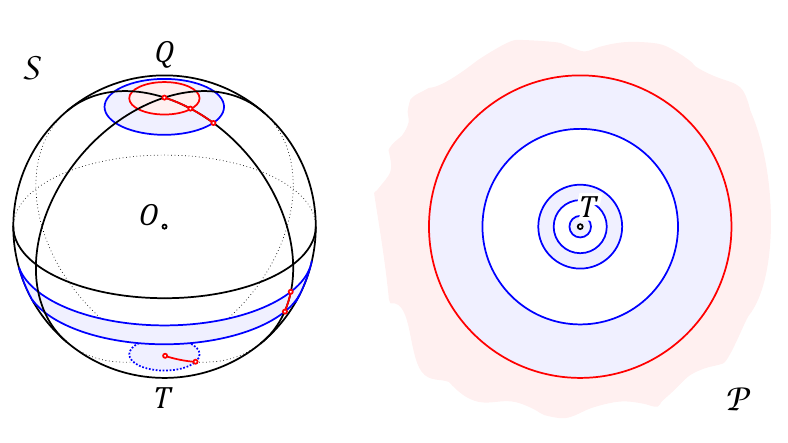}
\end{center}
\vspace*{-0.2cm}
\caption{Mapping of the compact surface of $\mathcal{S}$ onto the infinite Euclidean space $\mathcal{P}$. The great circle arcs drawn in red are all equal, consequently the shaded small circles and rings have the same height. Note that the pole $Q$ is excluded from the projection and the red circle encompassing it is mapped to an infinite volume outside the red circle in $\mathcal{P}$. The radii of the shells in $\mathcal{P}$ depend on the compactification map and are not to scale. We refer to the volume inside the red circle in $\mathcal{P}$ as the \textit{simulation volume}.}
\label{fig:1Dmap}
\end{figure}

Once an appropriate compactification map is selected, one can easily define an equal volume regular grid in the compact space $\mathcal{S}$. The grid should be aligned such a way, that the respective grid elements are centred on the poles $T$ and $Q$ of $\mathcal{S}$. For practical reasons, we use stereographic projection, equal spacing in the $\omega$ coordinate and equal-area HEALPix tiling \citep{2005ApJ...622..759G} in $\vartheta$ and $\varphi$. In the original three-dimensional space $\mathcal{P}$, when working in spherical coordinates, the equal volume regular grid on $\mathcal{S}$ will become a grid with elements of equal solid angle but of smoothly growing volume with $r$. The single grid element around the pole $Q$ opposite to the tangent point $T$ will correspond to an infinite volume region in the original three-dimensional space $\mathcal{P}$ that surrounds everything and stretches to infinity. It is plausible to assume $\delta = 0$ for this outermost element. All other grid elements together will correspond to a finite spherical region centred on the tangent point $T$, see~Fig.~\ref{fig:1Dmap}.

After the compactifying transformation and the grid are defined, the next step is to generate realistic initial conditions with a large enough three-dimensional volume so that all grid elements are populated except for the infinite element that contains the pole $Q$ of the compactified space $\mathcal{S}$. For this purpose, any existing initial condition generator can be used given that the output volume can be large enough and the particles show no periodicity. After the initial conditions are generated in the original, three-dimensional space, we map the particles into the compactified space and replace the individual particles within each grid element with a single, united particle that is placed at the centre of mass of the original particles falling into the grid element and carries the total mass and inertia. As a result, we get the initial conditions for the particles within a finite spherical simulation region of radius $R_\textnormal{sim}$. The resolution of the initial conditions is constant in the angular coordinates but decreases smoothly in the radial direction with growing distance from the center. The outlined method is probably the simplest way to construct a compactified simulation since the united particles can be developed forward with a regular $N$-body engine with the trivial $N^2$ Cartesian force calculators.

Particularly, for our simulations with stereographic compactification, we generated periodic initial conditions in a box of linear size $L_\textnormal{box}$ as input to the binning procedure. We chose a center in the periodic box for the tangent point, and projected the particle coordinates into the compactified space. The spherical cut to avoid the small region around the pole $Q$ was set to 
\begin{equation}
        \omega \leq \omega_{\text{max}} = 2 \cdot \arctan\left( \frac{L_{box}}{2\cdot R_S} \right)
\label{eq:ster_proj_rad_cut}
\end{equation}
in the compactified coordinate $\omega$, where $R_S$ was the radius of the compact sphere $\mathcal{S}$. With this cut, the radius of the simulation becomes $R_\textnormal{sim}=L_\textnormal{box}/2$.

\subsection{Isotropic boundary conditions in comoving coordinates}
\label{sec:boundary}

When generating the initial conditions with the binning method, as explained in the previous section, the grid element containing the pole of the compact spherical surface will correspond to the spherically symmetric, infinite volume of the Euclidean space that surrounds the entire simulation volume. As we pointed out earlier in Sec.~\ref{sec:static}, when static coordinates are used to run compactified simulations and no cosmological constant is present, the boundary conditions reduce to zero which means this infinite volume contributes zero force due to the spherical shell theorem. 

Let us now consider non-periodic simulations in comoving coordinates. It easily follows from Eq.~\ref{eq:Comoving_newtonian} that, if the boundary conditions are reduced to zero, a more or less homogeneous sphere of particles without significant peculiar initial velocities would collapse. To simulate a finite spherical volume of radius $R_\textnormal{sim}$ of the infinite universe in comoving coordinates, the effect of the mass outside the simulation volume must be taken into account and transformed into a non-zero boundary condition. When a compactification map is used, the resolution of the simulation decreases radially which in turn causes the amplitude $\delta$ of the density fluctuations in the grid cells to decrease. Consequently, the grid cell around the pole of the projection sphere can be approximated to have $\delta = 0$ and its contribution to the force can provide the necessary boundary conditions.

\begin{figure}
\begin{center}
\includegraphics[width=\columnwidth]{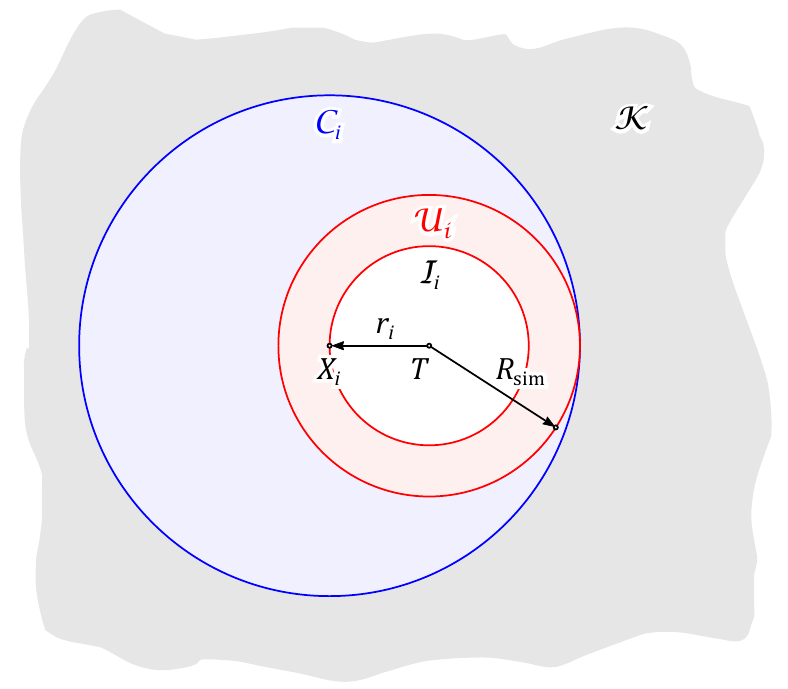}
\end{center}
\vspace*{-0.2cm}
\caption{Calculating the force on particle $i$ originating from the homogeneous infinite volume outside the simulation sphere. $\mathcal{K}$ is the infinite volume outside $R_\textnormal{sim}$, $\mathcal{I}_i$ is a spherical volume between $0 \leq r < r_i$ centred on the origin, where $r_i$ is the distance of particle $i$ from the origin. $\mathcal{U}_i$ is the simulation volume outside the radius of the particle. $\mathcal{C}_i$ is the minimal spherical volume centred on particle $i$ that encompasses the entire simulation. The shell theorem is applied to $\mathcal{K} - \mathcal{C}_i$ and $\mathcal{U}_i$ to calculate the force from $\mathcal{C}_i' = \mathcal{C}_i \setminus \left( \mathcal{U}_i \cup \mathcal{I}_i \right)$. See text for details.}
\label{fig:BoundaryForce}
\end{figure}

We calculate the force that particle $i$ feels from outside of the simulation volume using the shell theorem. Let us first consider exact homogeneity both inside and outside the simulation volume. Let $R_\textnormal{sim}$ be the radius of the simulation sphere and $r_i$ the distance of the particle from the centre of the simulation $T$, see Fig.~\ref{fig:BoundaryForce}. We divide the model universe into three distinct domains: $\mathcal{K}$ is the infinite volume outside the simulation sphere, $\mathcal{U}_i$ contains all shells of the simulation centred around $\mathcal{T}$ with a radius of $r > r_i$, and $\mathcal{I}_i$ is the volume inward from the particle where $r < r_i$. In the homogeneous case the force on particle $i$ is
\begin{equation}
0 = \mathbf{F}_{\mathcal{K}} + \mathbf{F}_{\mathcal{U}_i} + \mathbf{F}_{\mathcal{I}_i},
\label{eq:SumBoundaryForces}
\end{equation}
where $\mathbf{F}_{\mathcal{K}}$, $\mathbf{F}_{\mathcal{U}_i}$ and $\mathbf{F}_{\mathcal{I}_i}$ are the contributions originating from $\mathcal{K}$, $\mathcal{U}_i$, and $\mathcal{I}_i$, respectively. Let $\mathcal{C}_i$ be the spherical region centred on particle $i$ with a radius of $ R_\textnormal{sim} + r_i $, i.e. the minimal sphere centred on the particle that encompasses the entire simulation volume $\mathcal{U}_i \cup \mathcal{I}_i$, and $\mathcal{C}_i' = \mathcal{C}_i \setminus \left( \mathcal{U}_i \cup \mathcal{I}_i \right)$ see Fig.~\ref{fig:BoundaryForce}. The shell theorem says that the force $F_{\mathcal{U}_i}$ and the force from the region $\mathcal{K} \setminus \mathcal{C}_i$ are zero. Consequently, the force originating from $\mathcal{C}_i'$ must be
\begin{equation}
\mathbf{F}_{\mathcal{C}_i'} = - \mathbf{F}_{\mathcal{I}_i} = G m_i 4\pi\frac{\mathbf{x}_i}{|\mathbf{x}_i|}\int_{0}^{r_i} r^2\overline{\rho} dr = \frac{4 \pi G m_i}{3}\overline{\rho}\mathbf{x}_i,
\label{eq:CiForce}
\end{equation}
where $\overline{\rho}$ is the mean density.
If we drop the assumption of homogeneity inside the simulation radius $R_\textnormal{sim}$, the force from the region $\mathcal{U}_i \cup \mathcal{I}_i$ can be calculated with the right side of Eq.~\ref{eq:Comoving_newtonian}. Since we assume that the density field is always homogeneous outside $R_\textnormal{sim}$, the force from the region $\mathcal{C}_i$ and $\mathcal{K}$ must remain the same. Hence, the equations of motion in spherical simulations with isotropic boundary conditions are
\begin{equation}
\ddot{\mathbf{x}}_i = \sum\limits_{j=1; j \neq i}^{N} \frac{m_j\mathbf{F}(\mathbf{x}_i-\mathbf{x}_j, h_i+h_j)}{a(t)^{3}} - 2 \cdot  \frac{\dot{a}(t)}{a(t)} \cdot \dot{\mathbf{x}}_i + \frac{4 \pi G }{3}\overline{\rho}\mathbf{x}_i,
\label{eq:Comoving_Spherical_newtonian}
\end{equation}
where $\mathbf{F}(\mathbf{x}, h)$ is calculated using Eq.~\ref{eq:Force}.

\section{Newtonian simulations in static coordinates}
\label{sec:static}

With appropriately chosen initial and boundary conditions, Newtonian $N$-body simulations of an expanding universe are also possible in static coordinates. Initial conditions are typically generated in comoving coordinates, so in order to convert them to static coordinates and velocities, the following transformations are necessary:
\begin{eqnarray}
		\mathbf{X}_i &=& a \mathbf{x}_i
		\\
		\dot{\mathbf{X}}_i &=& \frac{\dot{a}}{a} \cdot \mathbf{X}_i + \dot{\mathbf{x}}_i,
	\label{eq:rescaling}
\end{eqnarray}
where $\mathbf{x}_i$ are the comoving, and $\mathbf{X}_i$ are the static coordinates, whereas $a$ is the time dependent scale factor taken at the time of the initial conditions. 

In the theoretical case of a CDM simulation with $\Lambda = 0$, the force law of Eq.~\ref{eq:Force} applies without any boundaries and the equations of motion in static coordinates become
\begin{equation}
	m_i\cdot\ddot{\mathbf{X}}_i = \sum\limits_{j=1; j \neq i}^{N} m_im_j\mathbf{F}(\mathbf{X}_i-\mathbf{x}_j, h_i+h_j).
\label{eq:Noncomoving_Newtonian}
\end{equation}
It is easy to see that these particles are moving away from an arbitrarily chosen origin and that the mass distribution around this centre must be isotropic in order to remain isotropic. This means that any finite volume simulation must start from an almost homogeneous sphere of particles centred on the origin with empty space beyond the simulation radius $R_\textnormal{sim}$. The simulation radius $R_\textnormal{sim}$ is limited by the requirement that velocities must be non-relativistic. As a consequence of the shell theorem, the equations of motion have no term from the boundary conditions, whereas the Hubble expansion is provided by the initial velocities of the particles. 

To conjure the boundary term originating from a cosmological constant that appears on the right-hand side of Eq.~\ref{eq:Noncomoving_Newtonian}, one should consider its effect as calculated from the Friedmann equations. In a universe with negligible matter and radiation the scale parameter grows as $a \propto \textnormal{e}^{H_0 t}$, consequently, the equation of motion in static coordinates should contain a term with exponential solution in the form of
\begin{equation}
	m_i\cdot\ddot{\mathbf{X}}_i = \sum\limits_{j=1; j \neq i}^{N} m_im_j\mathbf{F}(\mathbf{X}_i-\mathbf{x}_j, h_i+h_j) + H_0^2 \Omega_\Lambda \cdot m_i \cdot \mathbf{X}_i.
\label{eq:Noncomoving_Newtonian_Lambda}
\end{equation}

\section{Evaluating compactified simulations with angular power spectra}
\label{sec:angpowspec}

The most important immediate result of cosmological simulations is the evolution of the power spectrum. In case of compactified simulations, the radially decreasing resolution and the non-periodic boundary conditions complicate the estimation of the power spectrum via the spatial pair correlation function and its comparison to the power spectra of traditional periodic simulations. On the other hand, due to their single point of view nature and more similar geometry, compactified simulations might be better suited for comparison with observations, hence could become a useful tool in the design of cosmological surveys.

To evaluate compactified spherical simulations, computing the angular power spectrum in finite spherical shells appears to be a better approach than determining the spatial power spectrum, since the former is a better match to both, the simulations and observational geometry \citep{2011PhRvD..84f3505B}. Moreover, the process of determining the spatial power spectrum from observational data via the pair correlation function inherently assumes a background cosmology, which requirement could be alleviated by calculating the angular power spectrum instead in redshift bins.

The stereographic projection and a binning scheme based on HEALPix to generate the initial conditions, as it was outlined in Sec.~\ref{sec:stereo}~and~\ref{sec:initcond}, were also chosen to better match the requirement of calculating of the angular power spectrum. When comparing the outcome of compactified simulations to periodic ones, the same binning technique can be used within the periodic simulation box to compute the angular correlation function in radial shells that was used to compactify the initial conditions. When HEALPix is used for binning in the non-compactified coordinates, it is straightforward to use the \texttt{anafast} routine to compute the coefficients of the spherical harmonics. With HEALPix binning, the coefficients of the spherical harmonics are calculated as
\begin{equation}
\hat{a}_{lm}(r) = 
	\frac{4\pi}{N_\textnormal{pix}} 
	\sum\limits_{p=0}^{N_\textnormal{pix}-1} 
	Y^*_{lm}(\vartheta_p, \varphi_p)
	\left[\frac{\rho(r,\vartheta_p,\varphi_p)}{\overline{\rho}} - 1\right], 
\label{eq:alm}
\end{equation}
where $N_\textnormal{pix}$ is the number of the equal-area HEALPix cells that cover the entire surface of the shells and the variables $\vartheta_p$ and $\varphi_p$ go over the centres of the cells. The angular power spectrum is given by the usual
\begin{equation}
\hat{C}_l(r) = \frac{1}{2l+1}\sum\limits_m |\hat{a}_{lm}|^2,
\label{eq:angularpowerspectrum}
\end{equation}
summation over all directions.

\section{Results from the first S\lowercase{te}PS simulations}
\label{sec:results}

To compare the performance of StePS to a standard N-body code running on typical ``personal workstation'' hardware, we executed dark-matter-only $\Lambda$CDM simulations with our GPU-based implementation of the compactified universe method as well as with the GADGET-2 tree code. We chose GADGET-2 because it is freely available, and it is the most widely used code for cosmological simulations. We used \textit{Planck} best fit cosmology \citep{2016A&A...594A..13P} with $H_0 = 67.74$~km~s$^{-1}$~Mpc$^{-1}$, $\Omega_m = 0.3089$, $\Omega_\Lambda = 0.6911$, $\Omega_r = 0.0$ and $\sigma_8 = 0.8159$; the initial redshift was $z = 19$.

\subsection{Initial conditions}

In case of the GADGET-2 simulation the size of the periodic box was chosen to be $L_{box} = 1860.05$~Mpc. The initial conditions were generated using the 2LPTIC code \citep{2006MNRAS.373..369C}. The very same initial conditions were used as basis for the compactified StePS simulation: The radius of the four-dimensional sphere was chosen to be $R_s = 52.5$~Mpc. The diameter of this sphere determines the characteristic length on which the compactifying transformation is approximately linear. The $\omega \in [0, \omega_{\text{max}}]$ interval was divided uniformly into 224 elements. By leaving out the element containing the pole, the resulting simulation radius in the Euclidean space turned out to be $R_\textnormal{sim} = 930.03$~Mpc, and the tangent point of the stereoscopic transformation was set at $(x,y,z) = (1000\text{Mpc}, 1000\text{Mpc}, 930.03\text{Mpc})$. We used a HealPix grid of $N_\textnormal{side} = 32$ on the compactified spherical surface in the $\vartheta$ and $\varphi$ coordinates which corresponds to 12288 grid elements, each with an area of $1.73$~deg$^2$. This angular resolution is equivalent to a maximum multipole index of $l = 95$.

\begin{figure}
	\input{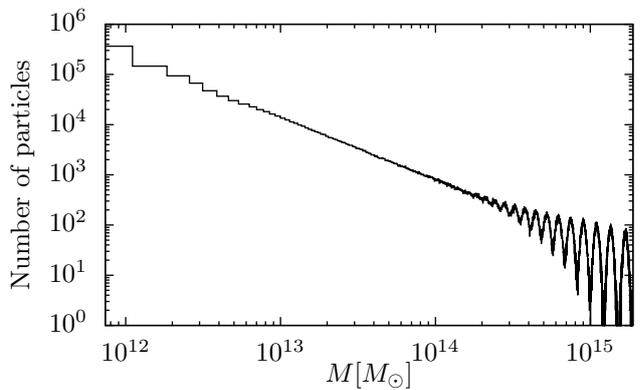}
\caption{Distribution of the particle mass in the compactified StePS simulation. See text for discussion.}
\label{fig:ParticleMasses}
\end{figure}

The distribution of particle masses is plotted in Fig.~\ref{fig:ParticleMasses}. Note, that the outermost elements of the compactifying grid are so large that the united particles are almost always positioned in the middle of the grid elements, hence the strange behaviour of the distribution at the high-mass end.

\subsection{Simulation hardware and performance}

Tab.~\ref{tab:simparams} summarizes the main parameters and computational performance of the two simulations. The GADGET-2 simulation was run on a dual-socket machine with two 10-core Intel Xeon E5-2630~v4 processors using 20 threads. The StePS simulation was performed on a single Nvidia GTX 1060 graphics card with 3~GB RAM. Although the comparison is not fair in the absolute sense since the periodic simulation had a fine resolution everywhere whereas StePS had  similar resolution only in a $105$~Mpc diameter sphere around the tangent point, it is clear that with more optimisation of the force calculation and with newer GPUs, a similar StePS simulations with reasonable resolution and excellent dynamic range could be performed in a few hours on a personal computer or even on a powerful laptop.

\begin{table}
\begin{tabular}{l | c | c}
\hline
					& GADGET-2 					& StePS \\ 
\hline
linear size $\left[ \textnormal{Mpc} \right]$
					& $L_\textnormal{box} = 1860$ & $D_\textnormal{sim} = 1860$ \\
simulated volume $\left[ \textnormal{Gpc}^3 \right]$
					& $6.44$ 					& $3.37$ \\
number of particles & $3.43 \times 10^8 $ 		& $1.47 \times 10^6$ \\
particle mass $\left[ M_\odot \right]$ 
					& $7.38 \times 10^{11}$		& $\sim 10^{12}$-$10^{15}$ \\
force calculation	& $\mathcal{O}(N \log N)$ 	& $\mathcal{O}(N^2)$ \\
memory use $\left[ \textnormal{MB} \right]$	
					& $\sim 40{,}000$ 				& $240$ \\
number of processor cores
					& $20$						& $1152$ \\
wall-clock time	$\left[ h \right]$	
					& $124$ 					& $39$ \\
\hline
\end{tabular}
\caption{Main parameters of the compared simulations.}
\label{tab:simparams}
\end{table}

%\begin{table}
%\begin{tabular}{| r | c |}
%$H_0$ [km/s/Mpc] & 67.74 \\
%$\Omega_m$ & 0.3089 \\
%$\Omega_\Lambda$ & 0.6911 \\
%$\Omega_r$ & 0.0 \\
%$\sigma_8$ & 0.8159 \\
%Initial redshift & 19.0 \\
%Linear size [Mpc]& 1860.05\\
%\end{tabular}
%\caption{Summary of cosmological and simulation input parameters. For the cosmological %parameters, we used the \textit{Planck} $\Lambda$CDM best-fit parameters %\citep{2016A&A...594A..13P}. The number of the particles was $1.47 \cdot 10^{6}$ in the %gadget simulation, and was $3.43 \cdot 10^{8}$ in the StePS simulation.}
%\label{tab:parameters}
%\end{table}

\subsection{Density distributions and angular correlation functions}

\begin{figure*}
    \centering
    \begin{subfigure}{6.5cm}

    \centering
    \includegraphics[width=\columnwidth]{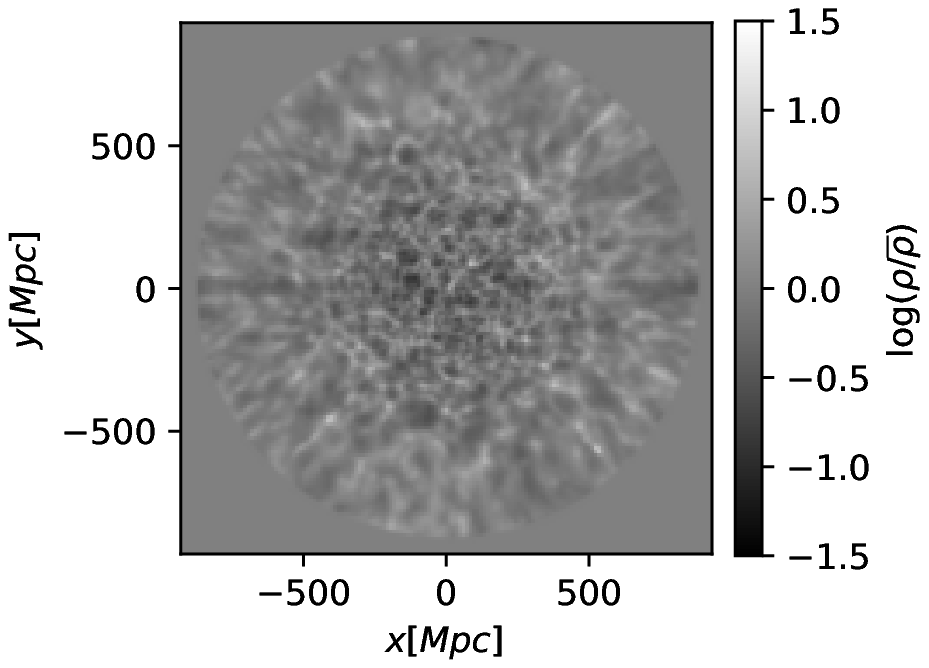}
    
    \end{subfigure}
    ~
    \begin{subfigure}{10cm}

	\vspace{-0.5cm}    
    \centering
    \includegraphics{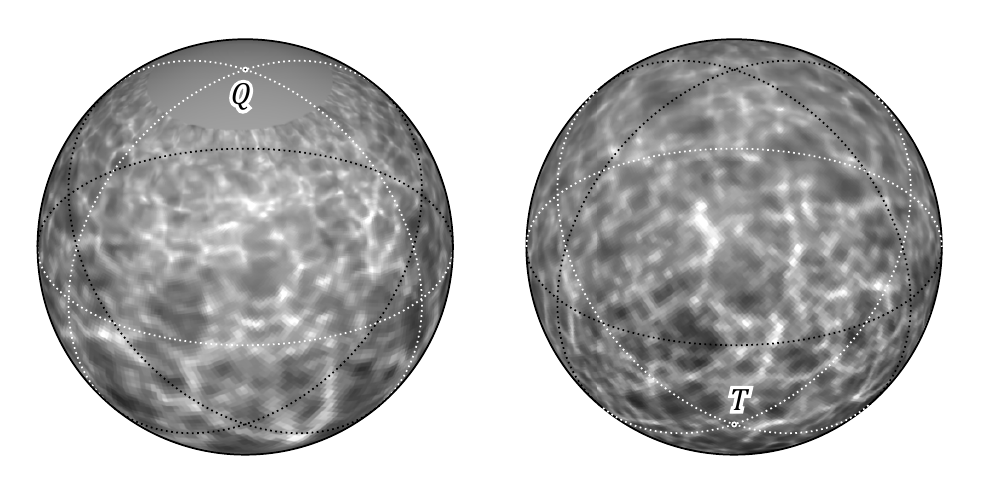}
    
    \end{subfigure}
    \caption{\textbf{Left:} The logarithmic density field of a particle distribution at $z=0$ as computed with the StePS method. The left panel shows a slice of the Euclidean space in which the equations of motion were evolved. The radially decreasing resolution is obvious. Outside the $R_\textnormal{sim} = 930$~Mpc simulation radius the isotropic boundary conditions of $\rho = \overline{\rho}$ was used. The density field was calculated from the point mass distribution with the the DTFE method \citep{2000A&A...363L..29S}. \textbf{Middle and right:} The spheres represent a $\Delta \vartheta$ shell of the StePS simulation in the $\omega, \vartheta, \varphi$ compactified coordinates from two different viewpoints. In the middle panel, the ``infinity'' element with $\delta = 0$ fluctuations is clearly visible around the projection point $Q$. In the right panel, we plot the same density field from a different viewpoint which shows the tangent point $T$ around which the simulation has the highest resolution. The fluctuations smoothly decrease from the tangent point $T$ toward the projection point $Q$ because equal area regions on the surface of the sphere correspond to increasingly larger volumes of the Euclidean space.}
    \label{fig:Output}
\end{figure*}

The distribution of the particles of the StePS simulation at redshift $z=0$ is plotted in Fig.~\ref{fig:Output} for a $\Delta z$ slice of the Euclidean space (left panel) and as a $\Delta \vartheta$ slice of the compactified space (right panel). The radially decreasing resolution is clearly visible on the former.

\begin{figure*}
    \centering
    \begin{subfigure}[b]{0.42\textwidth}
        \includegraphics[width=\textwidth]{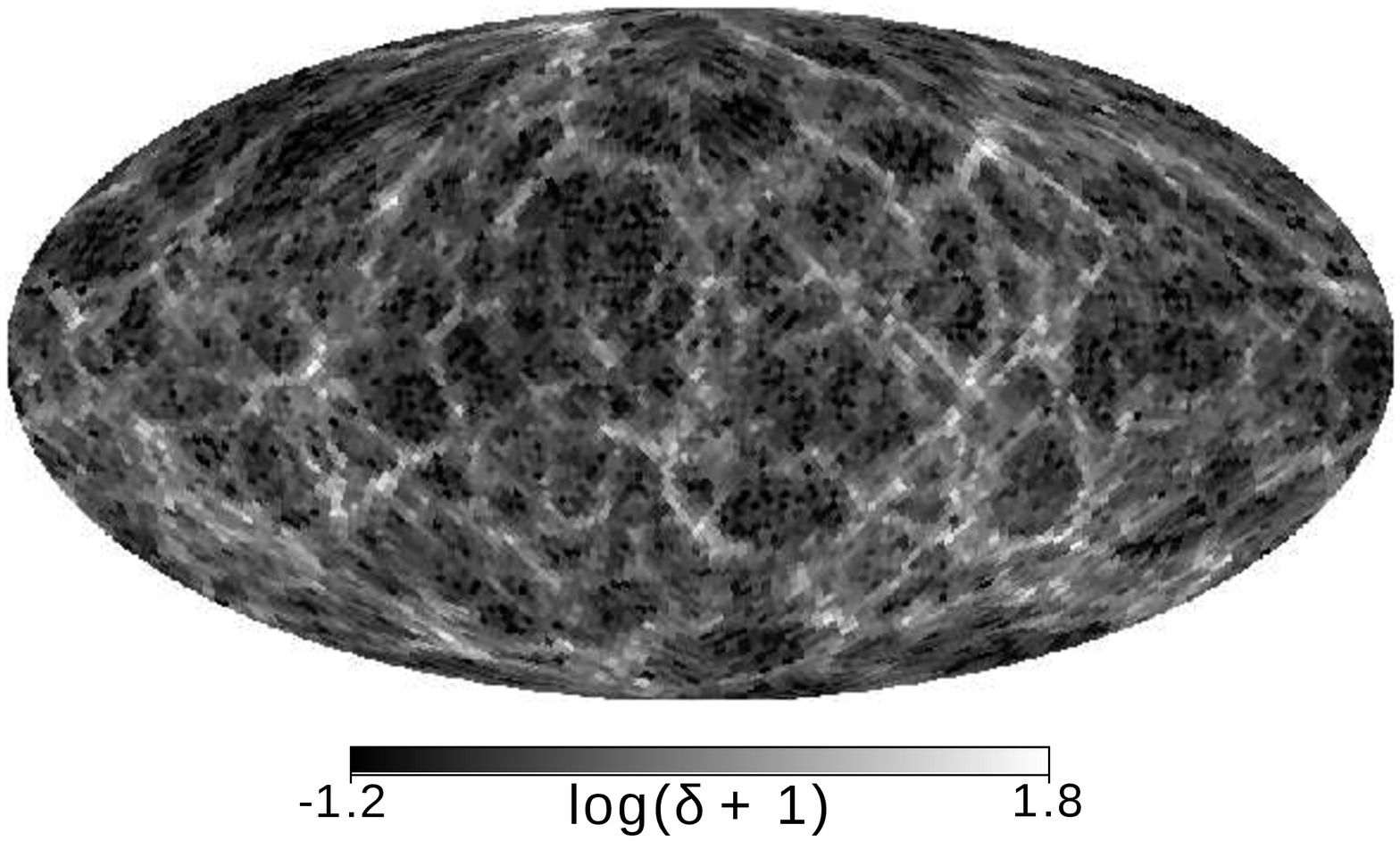}
        \caption{GADGET-2 simulation}
        \label{fig:GADGET_r100}
    \end{subfigure}
    ~ %add desired spacing between images, e. g. ~, \quad, \qquad, \hfill etc. 
      %(or a blank line to force the subfigure onto a new line)
    \begin{subfigure}[b]{0.42\textwidth}
        \includegraphics[width=\textwidth]{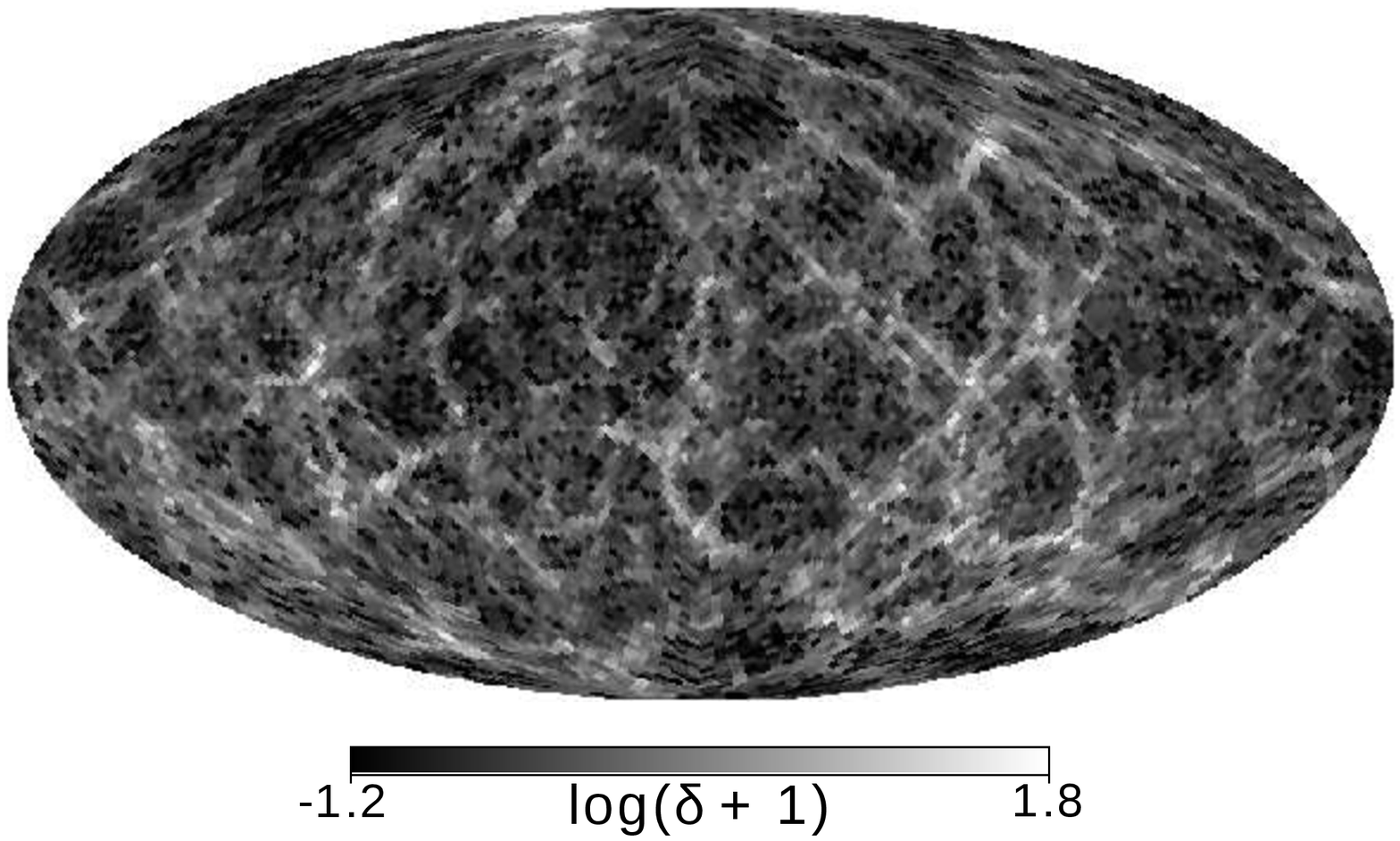}
        \caption{StePS simulation}
        \label{fig:SPCS_r100}
    \end{subfigure}
    \caption{The density fields in spherical shells with a radius of 100Mpc, calculated from the output of the periodic GADGET-2 simulation (left) and from our compactified StePS simulation (right). The plotted spherical shell of the periodic simulation was centred on the same point which was used as the origin of the compactified simulation.}
    \label{fig:shells}
\end{figure*}

\begin{figure}
    \centering
	\input{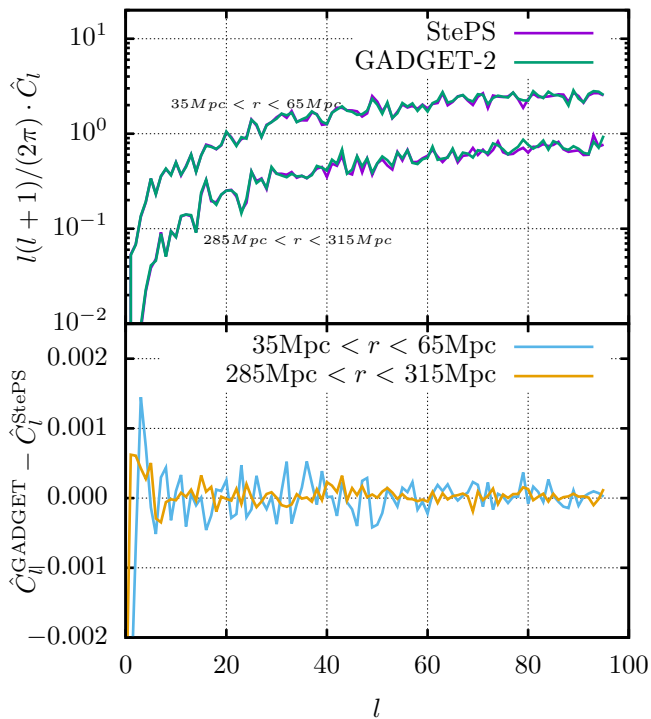}
    \caption{The $\hat{C}_l(r)$ angular power spectrum for the spherical shells of $35~\textnormal{Mpc} < r < 65~\textnormal{Mpc}$ and $285~\textnormal{Mpc} < r < 315~\textnormal{Mpc}$ centred on the same origin, as computed from the periodic GADGET-2 (green curve) and the StePS (purple curve) simulation. The lower panel shows the difference of the coefficients for each shell.}
    \label{fig:angpowspec}
\end{figure}

For the visual comparison of the two methods, in Fig.~\ref{fig:shells} we also plot the $z=0$ matter distribution for both GADGET-2 (left panel) and StePS (right panel) for a spherical shell of $R=100$~Mpc centred on the same Cartesian coordinates that were used as the tangent point to compactify the initial conditions. At this small radius from the origin, the mass and spatial resolution of the two simulations was approximately the same. As it can be clearly see in Fig.~\ref{fig:shells}, the resulting structure of the web of superclusters, filaments and voids is in complete agreement regardless of the simulation algorithm. To quantitatively compare the two methods, we plot the angular power spectra in Fig.~\ref{fig:angpowspec} for both simulations in two radial bins of $35~\textnormal{Mpc} < r < 65~\textnormal{Mpc}$ and $ 285~\textnormal{Mpc} < r < 315~\textnormal{Mpc}$. While the angular power spectra show significant noise due to the low resolution of the simulations, the agreement between the two is remarkable.

\section{Summary and future work}
\label{sec:summary}

We presented the fundamentals of a novel method to perform cosmological $N$-body simulations based on a compactification of the infinite universe and using isotropic boundary conditions instead of periodic ones. We demonstrated the feasibility of the method on easily accessible computer hardware using our implementation called StePS, which uses the three-dimensional stereographic projection for compactification. Within its range of validity, our simulation yields remarkably similar results to traditional $N$-body codes, such as GADGET-2. The key properties of our approach are the following:

\begin{itemize}

\item Unlike standard periodic simulations, our algorithm uses a strict $1/r^2$ Newtonian force law without any large scale modifications unsupported by observations. The softening of the force  on small scales is identical to classical simulation algorithms.

\item Our simulation, unlike the compact four-dimensional torus used in standard simulations, has an infinite topology, which is well supported by observations.

\item While standard simulations develop too many high-$k$ modes and too few low $k$ ones \citep{2011ApJ...737...11S}, our algorithm has a much better balance. Different compactifying maps can shift this balance at convenience.

\item Unlike traditional simulation cubes, the spherical simulation geometry matches observations. This motivates the use of angular power spectra in small $\Delta z$ bins of light cones. This corresponds to radical compression of the cosmological information without prior assumption of cosmological parameters \citep{2011PhRvD..84f3505B}. In contrast, to calculate the three-dimensional power spectra from observational data, a background cosmology must be assumed.

\item The simulation geometry is perfectly suited for creating light-cones with constant angular, albeit with varying radial and mass, resolution.

\item Our simulation predicts the power spectrum with unprecedented dynamic range for a given amount of memory and computational power. It is perfectly suited for modern graphics cards due to its low memory requirements, and no approximation to the force law is necessary. The standard $N^2$ algorithm is a lot faster for a given dynamic range than tree codes which makes our method perfect for the prediction needs of large surveys such as DES\citep{2017arXiv170609359K}, Euclid\citep{2011arXiv1110.3193L} or WFIRST \citep{2015arXiv150303757S}. 

\item Our simulation is perfectly suited for performing a series of simulations to estimate the power spectrum covariance matrix, correctly taking into account intra-survey and super-survey modes \citep{2015MNRAS.453..450C}.

\item Our simulation can run in either comoving or static coordinates. Comoving coordinates are adopted simply by adding a central force according to Eq.~\ref{eq:Comoving_Spherical_newtonian}. In case of static coordinates, dark energy can be simulated as a force proportional to the distance from the origin. 

\end{itemize}

Obviously, the current paper is limited only to introduce the key ideas of compactified cosmological simulations and to demonstrate their feasibility by presenting some preliminary results. Most importantly, rigorous future work is required to understand the systematics of the method. For instance, due to the covariance of low and high $k$ modes, the high $k$ modes estimated from the small volume high resolution region need correction.

Our method allows for optimising the degradation of spatial resolution with $r$ for specific purposes by using different compactification maps. For example, different maps could be ideal for maximising the dynamic range of the power spectrum or calculating precise mass functions. Perhaps different compactification maps should be used for vastly different surveys, such as the high redshift Euclid or low redshift surveys similar to SphereX \citep{2014arXiv1412.4872D}.

Although, to demonstrate the feasibility of our algorithm, we generated periodic initial conditions using a standard method in non-compactified Cartesian coordinates, an obvious step to take is to compute initial conditions directly in compactified coordinates, on the hypersurface of the four-dimensional sphere. Also, for sake of simplicity, we projected the compactified coordinates back to Cartesian and evolved the simulation using Euclidean distance in the force law. It might be possible to find a way to efficiently calculate forces directly in compactified coordinates which would allow running the simulations right on the surface of the four-dimensional sphere.

Given that the algorithm can simulate extreme volumes with relatively small computational resources, the super-horizon dynamics can be modified to include (linearized) general relativity. Also, since our simulation can run in static coordinates as well, instead of solving the Friedmann equations to provide the expansion background, the algorithm can compute the expansion of the universe directly. Dark energy can be taken into account as a force proportional to the distance from the simulation origin and no Friedmann equation is needed to rescale the coordinates. This approach will establish whether there is Newtonian backreaction, without neglecting shell crossing as previous studies \citep{1997A&A...320....1B, 2017arXiv170308809K} did.

\section{Acknowledgements}
This work was supported by NKFI NN 114560. IS acknowledges support from National Science Foundation (NSF) award 1616974. The authors would like to thank Robert Beck for stimulating discussions.
 
\bibliographystyle{mn2e}
\bibliography{SPCS}

\begin{thebibliography}{23}
\expandafter\ifx\csname natexlab\endcsname\relax\def\natexlab#1{#1}\fi

\bibitem[{{Bonvin} \& {Durrer}(2011)}]{2011PhRvD..84f3505B}
{Bonvin} C., {Durrer} R., 2011, \prd, 84, 063505

\bibitem[{{Buchert} \& {Ehlers}(1997)}]{1997A&A...320....1B}
{Buchert} T., {Ehlers} J., 1997, \aap, 320, 1

\bibitem[{{Carron}, {Wolk} \& {Szapudi}(2015){Carron}, {Wolk}, \&
  {Szapudi}}]{2015MNRAS.453..450C}
{Carron} J., {Wolk} M., {Szapudi} I., 2015, \mnras, 453, 450

\bibitem[{{Crocce}, {Pueblas} \& {Scoccimarro}(2006){Crocce}, {Pueblas}, \&
  {Scoccimarro}}]{2006MNRAS.373..369C}
{Crocce} M., {Pueblas} S., {Scoccimarro} R., 2006, \mnras, 373, 369

\bibitem[{{Dor{\'e}} {et~al}\mbox{.}(2014){Dor{\'e}}, {Bock}, {Ashby}, {Capak},
  {Cooray}, {de Putter}, {Eifler}, {Flagey}, {Gong}, {Habib}, {Heitmann},
  {Hirata}, {Jeong}, {Katti}, {Korngut}, {Krause}, {Lee}, {Masters},
  {Mauskopf}, {Melnick}, {Mennesson}, {Nguyen}, {{\"O}berg}, {Pullen},
  {Raccanelli}, {Smith}, {Song}, {Tolls}, {Unwin}, {Venumadhav}, {Viero},
  {Werner}, \& {Zemcov}}]{2014arXiv1412.4872D}
{Dor{\'e}} O. {et~al.}, 2014, ArXiv e-prints

\bibitem[{{G{\'o}rski} {et~al}\mbox{.}(2005){G{\'o}rski}, {Hivon}, {Banday},
  {Wandelt}, {Hansen}, {Reinecke}, \& {Bartelmann}}]{2005ApJ...622..759G}
{G{\'o}rski} K.~M., {Hivon} E., {Banday} A.~J., {Wandelt} B.~D., {Hansen}
  F.~K., {Reinecke} M., {Bartelmann} M., 2005, \apj, 622, 759

\bibitem[{{Hernquist}, {Bouchet} \& {Suto}(1991){Hernquist}, {Bouchet}, \&
  {Suto}}]{1991ApJS...75..231H}
{Hernquist} L., {Bouchet} F.~R., {Suto} Y., 1991, \apjs, 75, 231

\bibitem[{{Hockney} \& {Eastwood}(1988)}]{1988csup.book.....H}
{Hockney} R.~W., {Eastwood} J.~W., 1988, {Computer simulation using particles}

\bibitem[{{Kaiser}(2017)}]{2017arXiv170308809K}
{Kaiser} N., 2017, ArXiv e-prints

\bibitem[{{Krause} {et~al}\mbox{.}(2017){Krause}, {Eifler}, {Zuntz},
  {Friedrich}, {Troxel}, {Dodelson}, {Blazek}, {Secco}, {MacCrann}, {Baxter},
  {Chang}, {Chen}, {Crocce}, {DeRose}, {Ferte}, {Kokron}, {Lacasa}, {Miranda},
  {Omori}, {Porredon}, {Rosenfeld}, {Samuroff}, {Wang}, {Wechsler}, {Abbott},
  {Abdalla}, {Allam}, {Annis}, {Bechtol}, {Benoit-Levy}, {Bernstein}, {Brooks},
  {Burke}, {Capozzi}, {Carrasco Kind}, {Carretero}, {D'Andrea}, {da Costa},
  {Davis}, {DePoy}, {Desai}, {Diehl}, {Dietrich}, {Evrard}, {Flaugher},
  {Fosalba}, {Frieman}, {Garcia-Bellido}, {Gaztanaga}, {Giannantonio}, {Gruen},
  {Gruendl}, {Gschwend}, {Gutierrez}, {Honscheid}, {James}, {Jeltema}, {Kuehn},
  {Kuhlmann}, {Lahav}, {Lima}, {Maia}, {March}, {Marshall}, {Martini},
  {Menanteau}, {Miquel}, {Nichol}, {Plazas}, {Romer}, {Rykoff}, {Sanchez},
  {Scarpine}, {Schindler}, {Schubnell}, {Sevilla-Noarbe}, {Smith},
  {Soares-Santos}, {Sobreira}, {Suchyta}, {Swanson}, {Tarle}, {Tucker},
  {Vikram}, {Walker}, \& {Weller}}]{2017arXiv170609359K}
{Krause} E. {et~al.}, 2017, ArXiv e-prints

\bibitem[{{Lachieze-Rey} \& {Luminet}(1995)}]{1995PhR...254..135L}
{Lachieze-Rey} M., {Luminet} J., 1995, \physrep, 254, 135

\bibitem[{{Laureijs} {et~al}\mbox{.}(2011){Laureijs}, {Amiaux}, {Arduini},
  {Augu{\`e}res}, {Brinchmann}, {Cole}, {Cropper}, {Dabin}, {Duvet}, {Ealet},
  \& et~al.}]{2011arXiv1110.3193L}
{Laureijs} R. {et~al.}, 2011, ArXiv e-prints

\bibitem[{{Luminet}(2016)}]{2016Univ....2....1L}
{Luminet} J.-P., 2016, Universe, 2, 1

\bibitem[{{Monaghan} \& {Lattanzio}(1985)}]{1985A&A...149..135M}
{Monaghan} J.~J., {Lattanzio} J.~C., 1985, \aap, 149, 135

\bibitem[{{Navarro} \& {White}(1994)}]{1994MNRAS.267..401N}
{Navarro} J.~F., {White} S.~D.~M., 1994, \mnras, 267, 401

\bibitem[{{Norman} \& {Bryan}(1999)}]{1999ASSL..240...19N}
{Norman} M.~L., {Bryan} G.~L., 1999, in Astrophysics and Space Science Library,
  Vol. 240, Numerical Astrophysics, {Miyama} S.~M., {Tomisaka} K., {Hanawa} T.,
  eds., p.~19

\bibitem[{{O{\~n}orbe} {et~al}\mbox{.}(2014){O{\~n}orbe}, {Garrison-Kimmel},
  {Maller}, {Bullock}, {Rocha}, \& {Hahn}}]{2014MNRAS.437.1894O}
{O{\~n}orbe} J., {Garrison-Kimmel} S., {Maller} A.~H., {Bullock} J.~S., {Rocha}
  M., {Hahn} O., 2014, \mnras, 437, 1894

\bibitem[{{Peebles}(1980)}]{1980lssu.book.....P}
{Peebles} P.~J.~E., 1980, {The large-scale structure of the universe}

\bibitem[{{Planck Collaboration} {et~al}\mbox{.}(2016){Planck Collaboration},
  {Ade}, {Aghanim}, {Arnaud}, {Ashdown}, {Aumont}, {Baccigalupi}, {Banday},
  {Barreiro}, {Bartlett}, \& et~al.}]{2016A&A...594A..13P}
{Planck Collaboration} {et~al.}, 2016, \aap, 594, A13

\bibitem[{{Schaap} \& {van de Weygaert}(2000)}]{2000A&A...363L..29S}
{Schaap} W.~E., {van de Weygaert} R., 2000, \aap, 363, L29

\bibitem[{{Schneider} {et~al}\mbox{.}(2011){Schneider}, {Cole}, {Frenk}, \&
  {Szapudi}}]{2011ApJ...737...11S}
{Schneider} M.~D., {Cole} S., {Frenk} C.~S., {Szapudi} I., 2011, \apj, 737, 11

\bibitem[{{Spergel} {et~al}\mbox{.}(2015){Spergel}, {Gehrels}, {Baltay},
  {Bennett}, {Breckinridge}, {Donahue}, {Dressler}, {Gaudi}, {Greene}, {Guyon},
  {Hirata}, {Kalirai}, {Kasdin}, {Macintosh}, {Moos}, {Perlmutter}, {Postman},
  {Rauscher}, {Rhodes}, {Wang}, {Weinberg}, {Benford}, {Hudson}, {Jeong},
  {Mellier}, {Traub}, {Yamada}, {Capak}, {Colbert}, {Masters}, {Penny},
  {Savransky}, {Stern}, {Zimmerman}, {Barry}, {Bartusek}, {Carpenter}, {Cheng},
  {Content}, {Dekens}, {Demers}, {Grady}, {Jackson}, {Kuan}, {Kruk}, {Melton},
  {Nemati}, {Parvin}, {Poberezhskiy}, {Peddie}, {Ruffa}, {Wallace}, {Whipple},
  {Wollack}, \& {Zhao}}]{2015arXiv150303757S}
{Spergel} D. {et~al.}, 2015, ArXiv e-prints

\bibitem[{{Springel}(2005)}]{Gadget2_2005MNRAS.364.1105S}
{Springel} V., 2005, \mnras, 364, 1105

\end{thebibliography}

\end{document}